\documentclass[twocolumn,english,aps,prl,showpacs,superscriptaddress]{revtex4}
\usepackage{pslatex}
\usepackage[T1]{fontenc}
\usepackage[latin1]{inputenc}
\usepackage{amssymb}

\makeatletter

\providecommand{\LyX}{L\kern-.1667em\lower.25em\hbox{Y}\kern-.125emX\@}
\usepackage{babel}
\usepackage{graphicx}
\makeatother

\begin{document}

\title{Tuning the interactions of spin-polarized fermions using quasi-one-dimensional confinement }

\author{Brian E. Granger}
\affiliation{Institute for
Theoretical Atomic, Molecular and Optical Physics,
Harvard-Smithsonian CFA, Cambridge, MA 02138}
\email{bgranger@cfa.harvard.edu}
\author{D. Blume}
\affiliation{Department of Physics, Washington State
University, Pullman, WA 99164-2814}
\email{doerte@wsu.edu}

\begin{abstract}
The behavior of ultracold atomic gases depends crucially on the
two-body scattering properties of these systems. We develop a multichannel scattering 
theory for atom-atom collisions in
quasi-one-dimensional (quasi-1D) geometries such as atomic waveguides or
highly elongated traps. We apply our general framework to the low
energy scattering of two spin-polarized fermions and show that
tightly-confined fermions have infinitely strong interactions at a
particular value of the 3D, free-space $p$-wave
scattering volume. Moreover, we describe a mapping of this
strongly interacting system of two quasi-1D fermions
to a weakly interacting system of two 1D bosons.
\end{abstract}

%34.10.+x General theories and models of atomic and molecular collisions and interactions (including statistical theories, transition state, stochastic and trajectory models, etc.)
%34.50.-s Scattering of atoms and molecules
%03.75.-b Matter waves (for atom interferometry techniques, see 39.20.+q-in atomic and molecular physics)
%\draft

\maketitle

Two particle scattering is ubiquitous in physics. With
the achievement of quantum degeneracy in ultracold atomic gases 
\cite{Anderson1995a,DeMarco1999a}, renewed
interest in the scattering of two atoms at low temperatures has
arisen. This interest in atom-atom scattering stems from
the fact that, to first order, the many-body physics of an
ultracold atomic gas depends on a single atom-atom scattering
parameter \cite{Dalfovo1999a}. For spin-polarized bosons, this
parameter is the $s$-wave scattering length $a_{s}=-\lim
_{k\rightarrow 0}\tan \delta _{s}/k$, where $\delta _{s}$ is the
$s$-wave scattering phase shift at collision momentum $k$. For two
colliding spin-polarized fermions, in contrast, the antisymmetric
character of the wave function forbids $s$-wave scattering, and
instead, $p$-wave scattering becomes dominant. The interaction is
then characterized by the $p$-wave scattering volume $V_{p}=-\lim
_{k\rightarrow 0}\tan \delta _{p}/k^{3}$~\cite{beg-Suno2003a}.

An exciting feature of ultracold atomic gases is that the
effective two-body scattering properties can be manipulated by
external magnetic fields 
\cite{Inouye1998a,beg-Regal2003a},
or by strong confinement in one or more directions 
\cite{beg-Reichel1999a,beg-Muller1999a}. 
Of particular interest are quasi-1D geometries created by
atomic waveguides or optical lattices, which may, if loaded with ultracold
bosons
or fermions \cite{beg-OHara2002a,beg-Modugno2003a}, 
provide an opportunity to study novel 1D many
body states \cite{beg-Mattis1993a}, and to perform high precision measurements 
\cite{beg-Gustavson1997a}. 
Because the many-body physics of such quasi-1D systems
depends predominantly on the atom-atom scattering properties
\emph{in the waveguide}, it is imperative to understand how the
waveguide modifies the free space scattering properties.

The scattering of two bosons in a quasi-1D geometry has been
studied by Olshanii \cite{Olshanii1998a}. For a waveguide with
harmonic confinement the 3D Hamiltonian for the relative
coordinate $\vec{r}=(\rho,\phi,z)$ of two atoms of
mass $m$ reads:
\begin{equation}
H=-\frac{\hbar ^{2}}{2\mu }\nabla ^{2}+\frac{1}{2}\mu \omega
_{\perp }^2\rho ^{2}+V_{3D}(r),\label{eq:3DHam}\end{equation}
where $\mu =m/2$ denotes the reduced mass, and $V_{3D}(r)$ the
full atom-atom interaction potential. Using a regularized
zero-range potential, $V_{3D}(r)=\frac{2\pi }{\mu }\hbar
^{2}a_{s}\delta (\vec{r})\frac{\partial }{\partial r}r$, Olshanii
derives an effective 1D Hamiltonian,\begin{equation}
H_{1D}=-\frac{\hbar ^{2}}{2\mu }\frac{d^{2}}{dz^{2}}+g_{1D}\delta
(z)+\hbar \omega _{\perp },\label{eq:1Dham}\end{equation} and
coupling constant $g_{1D}$:\begin{equation} \frac{g_{1D}}{
a_{\perp}\hbar \omega_{\perp}}=2\frac{a_{s}}{a_{\perp
}}\left(1-\frac{a_{s}}{a_{\perp }}\left|\zeta (1/2)\right|
\right)^{-1},\label{eq:g1dbosons}\end{equation} which reproduce
the low energy scattering solutions of the full 3D Hamiltonian,
Eq.~(\ref{eq:3DHam}). This elegant result \cite{Olshanii1998a,beg-Bergeman2003a}
shows that the strong
confinement of the waveguide gives rise to an effective 1D
interaction, parameterized by $g_{1D}$, which can be tuned to
virtually any strength by changing the ratio of the 3D $s$-wave
scattering length $a_{s}$ to the transverse oscillator length
$a_{\perp }=\sqrt{\hbar /\mu \omega _{\perp }}$ through, for example,
application of an external magnetic field. Notably, the effective
1D interaction becomes infinitely strong ($g_{1D}$ diverges) when
the 3D scattering length takes on the particular \emph{finite}
value $a_{s}/a_{\perp }=1/\left|\zeta (1/2)\right|\approx 0.6848$,
where $\zeta(\cdot)$ is the Riemann zeta 
function~\cite{beg-Abramowitz}.

The results obtained for quasi-1D bosons lead us to the following
question: How do two spin-polarized \emph{fermions} under quasi-1D confinement
behave? To answer this question we develop a general framework
to obtain scattering solutions of the wave\-guide Hamiltonian,
Eq.~(\ref{eq:3DHam}), that are applicable to bosons \emph{and}
fermions at any energy, as well as to a large class of quasi-1D
scattering processes, single- \emph{or} multi-channel in nature,
at \emph{any} energy. When applied to identical fermions, this
framework shows that the confinement modifies the free-space
scattering properties of two spin-polarized fermions
significantly. In particular,
we show that two spin-polarized fermions --- just as two
spin-polarized bosons
--- can have infinitely strong effective 
interactions  for a {\em{finite}} 3D scattering
volume $V_p$ when
confined to a quasi-1D geometry.

A key quantity in our formalism is the familiar $K$-matrix
$\underline{K}^{3D}(E)$~\cite{beg-Aymar1996a}, 
which encapsulates the free-space (no confinement)
scattering physics of a
given atom-atom potential $V_{3D}(r)$ at all energies. One
advantage of using $K$-matrices is that the complications
 of using zero-range potentials are
avoided. For general multichannel collisions the 3D $K$-matrix,
$\underline{K}^{3D}$, can be determined accurately by a number of
numerical techniques \cite{beg-Julienne1997a,beg-Bohn1999a,beg-Kempen2002a}; for a single channel scattering
process it reduces to a diagonal matrix of scattering phase shifts
$\tan \delta _{l}$. Below, we derive an effective 1D $K$-matrix,
$\underline{K}^{1D}$, that i) is written in terms of the known 3D
$K$-matrix, $\underline{K}^{3D}$, and ii) fully characterizes the
effective interaction of two tightly-confined atoms.

For now, we assume an interaction potential $V_{3D}(r)$
such that the 3D K-matrix reaches its asymptotic limit at a distance
$r_c$ much smaller than the transverse confinement length $a_{\perp}$ ($r_c \ll a_{\perp}$).
While this
condition may not necessarily be fulfilled for a realistic atom-atom 
potential, one can often construct a model potential that
fulfills this requirement while mimicking the behavior of a
realistic two-body potential as closely as possible.
Configuration space can then be partitioned into two regions: i)
$r<r_{c}$. The two-body potential dominates, while the
confining potential is negligible. ii) $r > r_{c}$. The confinement
is felt, while the atom-atom potential is negligible. For $r
\approx r_c$, the full wavefunction can be conveniently written 
in spherical coordinates~\cite{beg-Aymar1996a}:
\begin{equation} \Psi _{\beta }(\mathbf{r})=\sum
_{l}F_{lm}(\mathbf{r})\delta _{l\beta
}-G_{lm}(\mathbf{r})K_{l\beta
}^{3D},\label{eq:spheresoln}\end{equation} 
where the 3D $K$-matrix,
$\underline{K}^{3D}$, contains all the ``scattering information'' 
about the atom-atom potential
$V_{3D}(r)$.
The energy normalized
regular solution $F_{lm}(\mathbf{r})$ at energy $E/\hbar \omega
_{\perp }=(ka_{\perp })^{2}/2$ is given in terms of the spherical
Bessel function $j_{l}(kr)$:\begin{equation}
F_{lm}(\mathbf{r})=Y_{lm}(\theta ,\phi )\frac{1}{\sqrt{\hbar
\omega _{\perp }a_{\perp }^3}}\sqrt{\frac{2ka_{\perp }}{\pi
}}j_{l}(kr).\label{eq:flr}\end{equation} The irregular solution
$G_{lm}(\mathbf{r})$ is identical to $F_{lm}(\mathbf{r}$), but
with $j_{l}(kr)$ replaced by $n_{l}(kr)$. In the spherical
representation, the parity of the relative wavefunction is simply
$\pi _{tot}=(-1)^{l}.$

To obtain the 1D scattering properties, the spherical solution,
Eq.~(\ref{eq:spheresoln}), must be propagated outward to large
$\left|z\right|$ where the cylindrical symmetry of the harmonic
confinement dominates. For $\left|z\right|\gg a_{\perp }\gg
r_{c}$, the wavefunction is a product of 2D harmonic oscillator
wavefunctions $\Phi _{nm}(\rho ,\phi )$ and free particle
solutions in $z$. We choose the energy normalized regular $\psi
_{nm}(\mathbf{r})$ and irregular $\chi _{nm}(\mathbf{r})$
cylindrical solutions 
to be eigenstates of the parity $\pi
_{tot}=\pi _{z}(-1)^{m}:$\begin{equation} \left\{ \begin{array}{c}
 \psi _{nm}(\mathbf{r})\\
 \chi _{nm}(\mathbf{r})\end{array}
\right\} =\Phi _{nm}(\rho ,\phi )\times \left\{ \begin{array}{c}
 f_{n}(z)\\
 g_{n}(z)\end{array}
\right\} .\label{eq:psinr}\end{equation}
 Here, the cylindrical channels are labeled by the harmonic oscillator
quantum number $n$ and the angular momentum $m$, so that the
$z$-direction momentum $q_{n}$ in each channel is defined by the
relationship,\begin{equation} \frac{E}{\hbar \omega _{\perp
}}=\frac{\left(ka_{\perp
}\right)^{2}}{2}=(2n+1+\left|m\right|)+\frac{\left(q_{n}a_{\perp
}\right)^{2}}{2}.\label{eq:etotal}\end{equation} The parity $\pi
_{z}$ is determined by the forms of the regular $f_{n}(z)$ and
irregular $g_{n}(z)$ free particle wavefunctions. 
To always work with
eigenstates of $\pi _{z}$ and to keep the regular and irregular functions
$90^{\circ }$ out of phase at both positive and negative
$z$, we use the conventions \cite{beg-Bianchi1994a},
\begin{equation}
f_{n}(z)=\frac{1}{\sqrt{\hbar \omega _{\perp }a_{\perp
}}}\frac{1}{\sqrt{\pi q_{n}a_{\perp }}}\times \left\{
\begin{array}{c}
 \cos (q_{n}\left|z\right|) \mbox{ for } \pi _{z}=1\\
 \frac{z}{\left|z\right|}\sin (q_{n}\left|z\right|) \mbox{ for }  \pi _{z}=-1\end{array}
\right.,\label{eq:fnz}\end{equation}
\begin{equation}
g_{n}(z)=\frac{1}{\sqrt{\hbar \omega _{\perp }a_{\perp }}}\frac{1}{\sqrt{\pi q_{n}a_{\perp }}}\times \left\{ \begin{array}{c}
 \sin (q_{n}\left|z\right|) \mbox{ for }  \pi _{z}=1\\
 -\frac{z}{\left|z\right|}\cos (q_{n}\left|z\right|) \mbox{ for }  \pi _{z}=-1\end{array}
\right. .\label{eq:gnz}\end{equation}
The full asymptotic wavefunction
at large $\left|z\right|$ can then be written in terms of the solutions
$\psi _{nm}(\mathbf{r})$ and $\chi _{nm}(\mathbf{r})$,
and a 1D $K$-matrix
$\underline{K}^{1D}$
\footnote{The 1D $K$-matrix depends on our conventions chosen for $f_n(z)$ and $g_n(z)$, which follow ref. \cite{beg-Bianchi1994a}.  Other conventions exist in the literature (see, for example, ref. \cite{beg-FanoRau1986a}).}:
\begin{equation} \Psi _{\alpha
}(\mathbf{r})=\sum _{n}\psi _{nm}(\mathbf{r})\delta _{n\alpha
}-\chi _{nm}(\mathbf{r})K_{n\alpha
}^{1D}.\label{eq:cylsoln}\end{equation}

To express the 1D $K$-matrix in terms of the 3D $K$-matrix we use
a frame transformation~\cite{beg-FanoRau1986a,beg-Fano1981a,beg-Harmin1982a}, 
which transforms one set of solutions of
the Schr\"odinger equation (labeled by a set of quantum numbers) to
another set of solutions (labeled by a different set of quantum
numbers). Such a transformation is exact and orthogonal when the
two sets of solutions satisfy the same Schr\"odinger equation
everywhere. A local, nonorthogonal frame 
transformation, on the other hand, is useful for transforming between two sets of
solutions that satisfy the same Schr\"odinger equation only in a
limited region of space. Following
Greene in his treatment of negative ion photodetachment in
magnetic fields~\cite{beg-Greene1987a}, 
we use the local frame transformation to relate
our spherical free particle solutions, Eq.~(\ref{eq:spheresoln}),
to our harmonically confined cylindrical solutions,
Eq.~(\ref{eq:cylsoln}), through a nonorthogonal matrix 
$\underline{U}$:
\begin{equation} \psi
_{nm}(\mathbf{r})=\sum _{l}F_{lm}(\mathbf{r})U_{ln} \,\, \mbox{
and } \,\, \chi_{nm}(\mathbf{r})=\sum
_{l}G_{lm}(\mathbf{r})\left(U^{T}\right)_{ln}^{-1}\label{eq:xinft}
.\label{eq:psinft}\end{equation}
The sum over $l$ in these expressions is understood to be over
$l=0,2,4,\ldots $ for $\pi _{z}=1$ and over $l=1,3,5,\ldots $ for
$\pi _{z}=-1$. The elements of the frame transformation matrix
$U_{ln}$ are calculated by projecting the expressions given
in Eq.~(\ref{eq:psinft}) onto the spherical 
harmonics~\cite{beg-Greene1987a}.
Using the transformation expressions, Eq.~(\ref{eq:psinft}), in Eq.~(\ref{eq:cylsoln}), 
the 1D $K$-matrix $\underline{K}^{1D}$ can be expressed in terms of
$\underline{K}^{3D}$ \emph{at all energies}:\begin{equation}
\underline{K}^{1D}=\underline{U}^{T}\underline{K}^{3D}\underline{U}.\label{eq:K1d3d}\end{equation}
This relationship is essentially exact for any
$\underline{K}^{3D}$, including multichannel cases, as long as
$r_{c}\ll a_{\perp }$.

Up to now, the cylindrical asymptotic wavefunction $\Psi _{\alpha
}(\mathbf{r})$, Eq.~(\ref{eq:cylsoln}), is written in terms of the
regular and irregular cylindrical 
functions solutions $\psi_{nm}(\mathbf{r})$ and $\chi _{nm}(\mathbf{r})$, 
which contain exponentially diverging pieces in channels $n$ that are
energetically closed ($E/\hbar \omega _{\perp
}<2n+1+\left|m\right|$). To obtain asymptotic solutions with the
correct, exponentially decaying boundary conditions in the closed
channels, these divergences must be eliminated. To do this, we use
the approach of multichannel quantum defect theory and partition
$\underline{K}^{1D}$ into a closed (``$c$'') and an open (``$o$'')
subspace~\cite{beg-Aymar1996a}:\begin{equation}
\underline{K}^{1D}=\left(\begin{array}{cc}
 \underline{K}_{oo}^{1D} & \underline{K}_{oc}^{1D}\\
 \underline{K}_{co}^{1D} & \underline{K}_{cc}^{1D}\end{array}
\right).\label{eq:K1dpartition}\end{equation}
Eliminating the closed channels results in a ``physical''
$K$-matrix in the open channels, $\underline{K}_{oo}^{1D,phys}$,
\begin{equation}
\underline{K}_{oo}^{1D,phys}=\underline{K}_{oo}^{1D}+i\underline{K}_{oc}^{1D}\left(\underline{1}-i\underline{K}_{cc}^{1D}\right)^{-1}\underline{K}_{co}^{1D}.\label{eq:Kphys}\end{equation}
The corresponding asymptotic wavefunction $\Psi _{\alpha }^{phys}$
--- $\alpha$ now denotes an open channel ---having the correct physical 
boundary conditions in all
channels involves a sum over only the open
channels:\begin{equation} \Psi _{\alpha }^{phys}(\mathbf{r})=\sum
_{n\in open}\Phi _{nm}(\rho ,\phi )\left[f_{n}(z)\delta
_{n\alpha }-g_{n}(z)K_{n\alpha
}^{1D,phys}\right].\label{eq:psiphys}\end{equation}
Equations~(\ref{eq:K1d3d}), (\ref{eq:Kphys}) and
(\ref{eq:psiphys})
provide a rigorous path from the full 3D scattering properties
encapsulated in $\underline{K}^{3D}(E)$, to an effective
1D system, Eq.~(\ref{eq:psiphys}), whose scattering
properties are given by $\underline{K}_{oo}^{1D,phys}(E)$. Our
framework shows that the closed channels contribute significantly
to the effective 1D scattering properties at energies
$E$ near closed channel resonances, where $\det
[\underline{1}-i\underline{K}_{cc}^{1D}\left(E\right)]
\approx 0$~(see alse~\cite{beg-Bergeman2003a}).

We now introduce an important simplification. When the 3D
scattering properties are dominated by the phase shift $\tan
\delta _{l}$ in a single partial wave $l$ ($K_{l^{\prime \prime
}l^{\prime }}^{3D}=\delta _{l^{\prime \prime }l^{\prime }}\delta
_{ll^{\prime }}\tan \delta _{l}$), the 1D $K$-matrix,
Eq.~(\ref{eq:K1d3d}), becomes a rank one matrix,\begin{equation}
K_{nn^{\prime }}^{1D}=\left(U^{T}\right)_{nl}\tan \delta
_{l}U_{ln^{\prime }}.\label{eq:Krankone}\end{equation}
 Thus, the closed channel part of $\underline{K}^{1D}$ also has rank
one with a single eigenvalue $\lambda _{c}$:\begin{equation}
\lambda _{c}={\mbox{Tr}}\underline{K}_{cc}^{1D}=\tan \delta
_{l}\sum _{n \in
closed}\left(U_{nl}\right)^{2}.\label{eq:lambdac}\end{equation}
This fact allows $\underline{K}_{oo}^{1D,phys}$,
Eq.~(\ref{eq:Kphys}), to be simplified by diagonalizing and
inverting the matrix
$\left[\underline{1}-i\underline{K}_{cc}^{1D}\right]$
analytically:\begin{equation}
\underline{K}_{oo}^{1D,phys}(E)=\underline{K}_{oo}^{1D}\left[1-i\lambda
_{c}(E)\right]^{-1}.\label{eq:Kphysrenorm}\end{equation}
Equation~(\ref{eq:Kphysrenorm}) shows that the ``bare''
1D $K$-matrix $\underline{K}_{oo}^{1D}$ is
renormalized by the closed channels physics encapsulated in the
eigenvalue $\lambda _{c}(E)$, which can be strongly energy
dependent. As the relative energy $E$ increases, new channels
become open and the dimension of $\underline{K}_{oo}^{1D,phys}$
 increases to reflect the
multichannel nature of the scattering. 

Application of our framework to the scattering of two identical 
bosons under strong transverse confinement 
 predicts the following general result:  when $a_s(E)/a_\perp=0.6848$, a divergence of the effective 1D interaction occurs at \emph{all} threshold energies, $E=\hbar \omega_\perp ( 2n+1 )$ for $m=0$, where a new transverse mode becomes open.  This extension of Olshanii's zero energy result \cite{Olshanii1998a} to all energies and multichannel cases suggests that a Tonks-like regime \cite{Olshanii1998a} can be realized experimentally in a ``high-temperature'' gas, $k_B T \gg \hbar \omega_\perp$.  Further details regarding our results for bosons will be published elsewhere~\cite{beg-Granger2003a}.

We now present an application of our framework to the low energy
scattering ($m=0$, $1<E/\hbar \omega_{\perp} <3$) of two spin-polarized fermions, whose 3D scattering properties are
parameterized by a single parameter,  the energy dependent
scattering volume $V_{p}(E)=-\tan \delta _{p}(E)/k^{3}$. 
In this case, the resulting effective fermionic 1D $K$-matrix,
Eq.~(\ref{eq:Kphysrenorm}), becomes:
\begin{eqnarray}
K^{1D,-} & \equiv & K^{1D,phys} = -\frac{6V_{p}}{a_{\perp }^{3}}q_{0}a_\perp \nonumber \\
 & \times  & \left[1-12\frac{V_{p}}{a_{\perp }^{3}}\zeta \left(-\frac{1}{2},\frac{1}{2}\left(3-\frac{E}{\hbar \omega _{\perp }}\right)\right)\right]^{-1}.\label{eq:Kphysfermions}
\end{eqnarray}
Here, the Hurwitz zeta function $\zeta(\cdot,\cdot)$~\cite{beg-Abramowitz}
arises from the eigenvalue $\lambda_c$ of
$\underline{K}_{cc}^{1D}$, $\lambda _{c}=
{\mbox{Tr}}\underline{K}_{cc}^{1D}$. The $K$-matrix given in
Eq.~(\ref{eq:Kphysfermions}) along with the odd-parity
wavefunction,\begin{equation} \psi ^{-}(z)\sim
\frac{z}{\left|z\right|}\left[\sin
\left(q_{0}\left|z\right|\right)+\cos
\left(q_{0}\left|z\right|\right)K_{00}^{1D,-}\right],\label{eq:psiminus}\end{equation}
provides a complete scattering solution to the waveguide
Hamiltonian, Eq.~(\ref{eq:3DHam}), when a single cylindrical
channel is energetically open.  The 1D $K$-matrix,
Eq.~(\ref{eq:Kphysfermions}), diverges when the scattering volume
$V_p$ has the particular value \begin{equation}
\frac{V_{p}^{crit}}{a_{\perp }^{3}}=\left[12\zeta
\left(-\frac{1}{2},\frac{1}{2}\left(3-\frac{E}{\hbar \omega
_{\perp
}}\right)\right)\right]^{-1}.\label{eq:vpcrit}\end{equation} This
implies that two spin-polarized quasi-1D fermions have infinitely
strong interactions for a finite 3D scattering volume $V_p=V_p^{crit}$.

Next, we derive an effective 1D Hamiltonian that describes many of
the low energy properties of two spin-polarized fermions in a waveguide [see Eq.~(\ref{eq:3DHam})]. Importantly, the 1D zero-range
potential $g_{1D}\delta (z)$ [see Eq.~(\ref{eq:1Dham})], which has
been very successful in treating bosons~\cite{Olshanii1998a,beg-Mattis1993a}, 
cannot be used {\em{directly}}
since it results in an unphysical scattering amplitude for fermions. 
One way around this difficulty would be to use a
zero-range potential that gives a meaningful scattering amplitude for
fermions~\cite{beg-Cheon1999a}. 
Alternatively, we propose to map the fermionic
${K}^{1D,-}$, Eq.~(\ref{eq:Kphysfermions}), to a bosonic 1D
K-matrix (along with the corresponding wavefunctions).
Mappings between fermions and bosons
are important in theoretical treatments of 1D {\em{many body}}
systems, as they allow one to understand systems of strongly
interacting 1D bosons (fermions) by mapping them to weakly
interacting systems of 1D fermions (bosons)~\cite{beg-Mattis1993a}.

At low energies, the 1D scattering wavefunction $\psi^-(z)$ for
two fermions is given by Eq.~(\ref{eq:psiminus}) while that for
two bosons reads:\begin{equation} \psi ^{+}(z)\sim \left[\cos
\left(q_{0}\left|z\right|\right)-\sin
\left(q_{0}\left|z\right|\right)K^{1D,+}\right],\label{eq:psiplus}\end{equation}
where $K^{1D,+}$ denotes the even parity 1D $K$-matrix. With the
choice,\begin{equation}
K_{00}^{1D,+}=-1/K_{00}^{1D,-},\label{eq:Kmapped}\end{equation}
the bosonic wavefunction can be written in terms of the fermionic
one (and vice versa): \begin{equation} \psi
^{+}(z)=\frac{\left|z\right|}{z}\psi
^{-}(z)/K_{00}^{1D,-}.\label{eq:psimapped}\end{equation}
Application of the proposed mapping, Eqs. (\ref{eq:Kmapped}) and
(\ref{eq:psimapped}), to our effective 1D $K$-matrix for two
fermions, Eq.~(\ref{eq:Kphysfermions}), results in an equivalent
system of two 1D bosons interacting through the potential
$g_{1D}^{map}\delta (z)$, with the ``mapped coupling
constant'':\begin{equation} \frac{g_{1D}^{map}}{a_{\perp }\hbar
\omega _{\perp }}=\frac{-a_{\perp
}^{3}}{6V_{p}}\left[1-12\frac{V_{p}}{a_{\perp }^{3}}\zeta
\left(-\frac{1}{2},\frac{1}{2}\left(3-\frac{E}{\hbar \omega
_{\perp }}\right)\right)\right].\label{eq:mappedg1d}\end{equation}
This remarkable result implies that two spin-polarized quasi-1D
fermions with {\em{infinitely strong}} interactions [see
Eq.~(\ref{eq:vpcrit})], can be mapped to a system of
\emph{non-interacting} bosons. More importantly, however, the mapped 
$g_{1D}^{map}$, Eq.~(\ref{eq:mappedg1d}), applies to any interaction strength $V_p$.  

\begin{figure}
 \includegraphics[width=3.25in,height=2.5in]{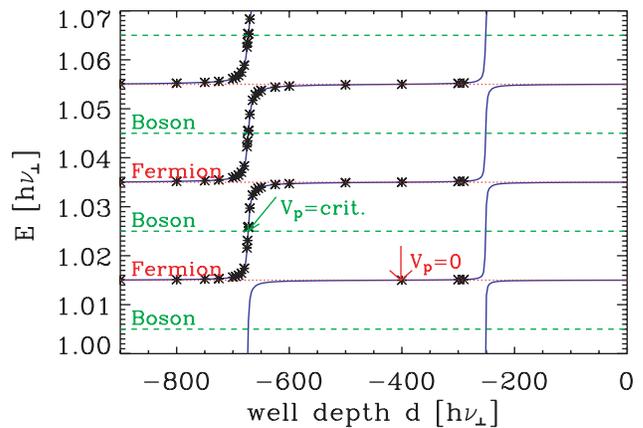}
\caption{
3D energy levels (solid lines) together with 1D energy levels (asterisks)
as a function of the two-body well depth $d$ ($b$ is fixed at
$b=0.13 a_{\perp}$) for two spin-polarized fermions under
highly-elongated confinement. Straight dotted
lines indicate the energy levels of two non-interacting spin-polarized fermions, while straight dashed lines indicate those for
two non-interacting bosons. Arrows indicate values of $d$
for which the corresponding scattering volume $V_p(E)$
has the particular values $V_p=0$ and $V_p=V_p^{crit}$, respectively.
}
\label{fig:spectrum}
\end{figure}

To confirm our analytical results we have performed numerical
calculations for two spin-polarized
fermions in a highly elongated trap interacting through
a two-body model potential, $V_{3D}(r)=d/\mbox{cosh}^2(r/b)$. 
We use a B-spline basis set to solve the 3D 
Schr\"odinger equation for the
Hamiltonian given in Eq.~(\ref{eq:3DHam}) with the confining
potential replaced by $\frac{1}{2}\mu \omega_{\perp}^2 (
\rho^2+\lambda^2 z^{2})$, where $\lambda=0.01$. The antisymmetry of
the fermionic wave functions is enforced by only including
odd-parity states. Solid lines in
Fig.~\ref{fig:spectrum} show the resulting energy spectrum, 
$E>\hbar \omega_{\perp}$,
as a function of the two-body well depth $d$. Straight dotted
lines indicate the energy levels of two non-interacting spin-polarized fermions, while straight dashed lines indicate those for
two non-interacting bosons. As predicted analytically
(and indicated by an arrow in Fig.~\ref{fig:spectrum}), the full 3D
energy levels coincide with that of two non-interacting bosons
for well depths $d$ corresponding to $V_p=V_p^{crit}$.

Figure~\ref{fig:spectrum} additionally shows the spectrum of the
1D bosonic Hamiltonian, Eq.~(\ref{eq:1Dham}), with the additional
potential $\frac{1}{2}\mu \lambda^{2}\omega_{\perp}^2z^{2}$, using the mapped
coupling constant $g_{1D}^{map}$ [Eq.~(\ref{eq:mappedg1d})]
(asterisks). To account for the energy-dependence of
$g_{1D}^{map}$ we self-consistently solve for the scattering
volume $V_p$ and the 1D energy levels for each well depth $d$ \cite{beg-Bolda2003a};
details will be published elsewhere \cite{beg-Granger2003a}. 
Figure~\ref{fig:spectrum}
shows that the 1D energies agree very well with the 3D
energies. 
Two
interacting spin-polarized fermions under quasi-1D confinement can
hence be mapped to a system of two 1D bosons interacting through a
$\delta $-function potential with a mapped coupling strength
$g_{1D}^{map}$, Eq.~(\ref{eq:mappedg1d}).
We note that present-day experiments \cite{beg-Regal2003a} can access regimes 
where our theory applies;
the question of how to experimentally
verify our predictions, however, is beyond the scope of this paper.

In summary, this Letter develops a general framework for treating
atom-atom scattering under quasi-1D
confinement.  The framework includes multichannel collisions at arbitrary energies
and is limited only in the assumption that the characteristic length of the
two-body potential is small compared to the confinement length $a_{\perp}$. Application
of the framework to two spin-polarized fermions in a waveguide
shows that the 3D scattering behavior, which is assumed to
be characterized completely by the scattering volume $V_p$, is
altered significantly due to the presence of the tight transverse
confinement. Specifically, two quasi-1D fermions can have
infinitely strong interactions for a finite 3D scattering volume.
Finally, the system of two quasi-1D fermions can be mapped to a
system of two 1D bosons.

Further study is required to understand the full implications of these results for the many body physics of quasi-1D spin-polarized 
fermions.  At this point, we only mention that any many body theory will have to carefully account for the energy-dependence of the mapped coupling constant $g_{1D}^{map}$.

Discussions with Chris Greene and Mike Moore were helpful in this
work. This work was supported by the NSF through a grant to ITAMP, 
Harvard-Smithsonian CFA.  DB acknowledges additional funding by the NSF under grant 0331529.

%\bibliographystyle{apsrev}
%\bibliography{1dscat,BEC}

\end{document}